# Two distinct logical types of network control in gene expression profiles

Carsten Marr[1‡], Marcel Geertz[2‡], Marc-Thorsten Hütt[1], Georgi Muskhelishvili[2*]

**1** Computational Systems Biology Group, **2** Molecular Genetics Group, Jacobs University, Campus Ring 1, 28759 Bremen, Germany

‡ These authors contributed equally to this work
* To whom correspondence should be addressed. E-mail: g.muskhelishvili@jacobs-university.de

**In unicellular organisms such as bacteria the same acquired mutations beneficial in one environment can be restrictive in another. However, evolving Escherichia coli populations demonstrate remarkable flexibility in adaptation. The mechanisms sustaining genetic flexibility remain unclear. In E. coli the transcriptional regulation of gene expression involves both dedicated regulators binding specific DNA sites with high affinity and also global regulators - abundant DNA architectural proteins of the bacterial chromoid binding multiple low affinity sites and thus modulating the superhelical density of DNA. The first form of transcriptional regulation is dominantly pairwise and specific, representing digitial control, while the second form is (in strength and distribution) continuous, representing analog control. Here we look at the properties of effective networks derived from significant gene expression changes under variation of the two forms of control and find that upon limitations of one type of control (caused e.g. by mutation of a global DNA architectural factor) the other type can compensate for compromised regulation. Mutations of global regulators significantly enhance the digital control; in the presence of global DNA architectural proteins regulation is mostly of the analog type, coupling spatially neighboring genomic loci; together our data suggest that two logically distinct – digital and analog – types of control are balancing each other. By revealing two distinct logical types of control, our approach provides basic insights into both the organizational principles of transcriptional regulation and the mechanisms buffering genetic flexibility. We**

**anticipate that the general concept of distinguishing logical types of control will apply to many complex biological networks.**

**Introduction**

Understanding the organizational logic of transcriptional regulation is central for deciphering those biological processes in which the involvement of genetic control is essential. For the classical model organism E. coli the largest electronically accessible network integrating the data on the transcriptional regulation of genes is available [1]. The interlinked elements form a complex structure, which is essentially of digital nature (digital refers here to the fact that the network provides static information on the connections between unique, discontinuous components [2], e.g. a particular pair of regulating and regulated gene). Notably, such pair-wise connections are not necessarily reflected in genomic expression profiles [3,4] indicating that not all the interactions given in the network occur at all times. Furthermore, this type of network does not account for the analog mode of gene regulation via alterations of DNA topology – a long known control mechanism revived by recent DNA microarray analyses [5-7] (analog refers here to the fact that the expression of specific genes is under the control of continuous information provided by spatial distributions of supercoiling energy in the genome [8]). Indeed, transcriptional responses to alterations of DNA superhelicity reveal non-trivial spatial patterns, raising new questions on the coordination of genomic transcription [6,8]. The interplay between chromosomal organization and patterns in gene expression has been in the focus of computational analyses recently [9,10]. Thus, a holistic theory of transcriptional regulation has to include the relationships between these two logically distinct types of information and therefore has to distinguish them in the first place. From these considerations for understanding organizational principles of transcriptional regulation we assume a working model in which the impact of two distinct logical types of control – one of digital and another of analog type – are to be clearly distinguished and related to each other.

In the following, we will translate the patterns in gene expression changes observed under systematic variation of the two types of control into effective networks and study their connectivity. The effective networks are derived as subnetworks of two larger (static) networks: (1) the transcriptional regulatory network based upon the action of dedicated transcription factors; (2) spatial proximity of two genes on the circular chromosome.

We will statistically compare the properties of these effective networks with those from random variants obtained by randomly sampling the static networks with a certain number of expression changes. The core quantity derived from this comparison is the ratio of connected to isolated nodes (control ratio) and, furthermore, its z-score with respect to the random networks. This z-score we denote the confidence level of the particular control type (control type confidence, CTC).

**Results**

In this study we aim at understanding the relationships between these two distinct types of control in transcriptional regulation by using the model system of exponentially growing *E. coli* cells. The rationale is to investigate transcript profiles obtained under conditions where we either modulate the analog component of regulation under constant digital control, or modulate the digital component keeping the analog control constant. We modulate the analog component by experimentally varying the negative superhelical density ($-\sigma$) of chromosomal DNA within the same "constant" genetic background. Such variation of $-\sigma$ is carried out within three genetic backgrounds – the wild type and two mutant *E. coli* strains lacking one of the two abundant DNA architectural proteins, either FIS or H-NS. These comparisons produce the so-called intra-strain transcript profiles [8] (see Figure 1). Modulation of the digital component is achieved by mutating genes of the same two global DNA architectural proteins (either *fis* or *hns*) and comparing the wild type and mutant transcript profiles at a single constant superhelical density – either DNA relaxation ($-\sigma < 0.033$) or high negative supercoiling ($-\sigma > 0.08$). These comparisons produce the so-called inter-strain transcript profiles [8]. The first approach enables us to assess the impact of digital control in transcriptional regulation under variation of the analog component. The second approach allows us to assess the impact of analog control under variation of the digital component. We thus obtain seven data sets: three distinct intra-strain transcript profiles reflecting digital-type control (wt, *fis*, *hns* for wild type, *fis* mutant and *hns* mutant backgrounds respectively), and four inter-strain profiles (wt-*fis* and wt-*hns* both at relaxation (↓σ) and high negative supercoiling (↑σ) reflecting analog-type control (Figure 1).

The transcriptional regulatory network (TRN) of *E. coli* is the basis of many recent studies on network architecture [11,12], as well as on the consistency of the network with expression profiles [3,4]. To assess the impact of digital-type control we analyze subnets of the TRN of *E. coli* spanned by genes with significantly changed expression in our three intra-strain transcript profiles, the *effective* TRNs (Figure 2). A convenient way of formalizing properties of these subnets is to analyze the ratio of genes with and without links, respectively. We define the control ratio $R$ as the number of connected nodes divided by the number of isolated nodes in the effective TRN. Comparing this ratio with corresponding random models (see Figure 2) we obtain the z-score of this ratio, which we denote the control type confidence (CTC). The CTC quantifies how much above-random connectivity is found in the effective network and, consequently, how much control the network exerts on the expression profile. Formally, the digital CTC is the z-score of the control ratio $R$ for the effective TRN, when compared to the distribution of control ratios, where the same number of affected nodes is mapped randomly on the TRN. We find a ratio $R > 1$ and CTC values beyond 2 only for two data sets – the intra-strain profiles of the *fis* and *hns* mutants (Figure 3), indicating that compared to wild type, in both mutants transcriptional regulation comprises a large proportion of digital-type control. Thus paradoxically, mutations of global regulators, which represent hubs targeting disproportionately large numbers of genes in the TRN, increase rather than decrease the number of connected genes and thus enhance digital control in the effective TRNs obtained. At the same time, effective TRNs of the four inter-strain profiles did not deviate substantially from a random model (Figure 3), as expected from our experimental design. This is because in the intra-strain profiles the constant digital control (background-specific TRN) enables to measure its impact under the variation of analog component (superhelical density, σ), whereas in the inter-strain transcript profiles the TRN itself is a variable. The concept of an effective TRN thus allows quantifying the contribution of digital control to genomic expression patterns.

In Figure 4, we demonstrate schematically the difference between the digital (a) and the analog (b) type of control. In order to analyze the digital and analog types of control on the same methodological basis, we convert the chromosomal neighborhoods of genes into a network, denominated the gene proximity network (GPN) (see Materials and Methods for details on the construction algorithm). The GPN subnet analysis of the inter-strain transcript profiles exposes the extent of spatial connectivity between the neighboring loci and reveals if significant expression changes are clustered on the genome. Alike the digital CTC, the analog CTC represents the z-score of the control ratio for the effective GPN, when compared to the distribution of control ratios from a null model, where the same number of affected nodes is mapped randomly on the genome. An important difference between the FIS and H-NS effects is of note here. Although both are abundant DNA binding proteins occupying multiple chromosomal sites, H-NS acts as universal repressor for the bacterial genome, whereas FIS is implicated in organization of superhelical loops and activation of genes involved in metabolism and growth [8,13-15]. Thus, directionally opposite effects – one largely of activation and another of global repression – are expected to underlie the GPNs in the inter-strain comparisons of wild type strain with *fis* and *hns* mutants, respectively. A schematic view on the different effects of FIS, which activates transcription by stabilizing branch points in DNA, and H-NS, which represses transcription by stabilizing tightly interwound DNA plectonemes, is shown in Figure 4. The GPNs of the *hns* mutant primarily reflect the spatial connectivity between de-repressed genetic loci, especially since H-NS represses whole regulatory systems rather than selectively targeted individual gene components [14,16]. We therefore assign to the wild type background the genes with positive log ratio in both *fis* experiments (wt-*fis* ↓σ and ↑σ)) and the genes with negative log ratio in both *hns* experiments (wt-*hns* ↓σ and ↑σ). We set the GPN threshold parameter (see Materials and Methods for a detailed explanation of the GPN construction) to 5 kbp to approximate the size limits of previously reported topological domains in the *E. coli* chromosome (17, 18). However, a consistent difference of calculated CTCs is observed over the whole sensible range of GPN thresholds (Figure 5c). In both inter-strain GPNs derived from the comparisons of wild type with *fis* mutant (wt-*fis* ↓σ and ↑σ) in Figure 5a and 5b, the genes with a positive log ratio exhibit a higher CTC. As expected, an opposite result is obtained with *hns* mutant (wt-*hns* ↓σ and ↑σ) in Figure 5a and 5b, where genes with a negative log ratio clearly exhibit a higher CTC. The GPN analysis of none of the three intra-strain profiles deviate strongly from the random model, because due to experimental design in these transcript profiles the analog component (i.e. the superhelical density –σ) itself is a variable (data not shown). Our GPN analyses thus indicate a high spatial connectivity of neighboring genes in wild type strains, comprising both the activating effect of FIS and the global repressing effect of H-NS (Figure 5). Analysis of an operon based proximity network does not substantially alter the observed results (data not shown). We infer that the abundant bacterial chromoid proteins FIS and H-NS substantially contribute to the analog-type of transcriptional control employing the spatial connectivity between neighboring genetic loci.

**Discussion**

Spatial organization of transcription has been observed in both prokaryotes and eukaryotes [19,20]. In *E. coli* this phenomenon can be readily rationalized on the basis of topological domains of variable size underlying the organization of bacterial chromosome [17,18,21]. Furthermore, the observed preponderance of analog-type control in wild type cells (see Figure 6) is in keeping with the property of chromatin proteins to stabilize supercoils and modulate the distributions of effective superhelicity in genome [13,15]. Finally, our finding that the spatial connectivity is substantially altered by mutations of *fis* and *hns* genes is fully consistent with recent study identifying both *fis* and *hns* among the few genes affecting the formation of topological barriers in the *E. coli* chromosome [22].

In summary, we present a generic approach allowing both, to distinguish and to assess the relationships between two logically distinct types of transcriptional control. This approach is essential for understanding the basic organizational principles of transcriptional regulation, especially since organizationally distinct architectures of sub-networks have been recently described in eukaryotes [23]. Using this approach we demonstrate that variation of the analog component of regulation (DNA superhelicity) effectively exposes the contribution of digital-type control (represented by the TRN) to transcriptional regulation, which is significantly increased in *E. coli* strains lacking global DNA architectural proteins. In turn, alterations of the digital component (deletions of genes encoding global DNA architectural proteins) expose a substantial contribution of analog-type control (approximated by the GPN) to transcriptional regulation in wild type cells. Taken together our data suggest that two logically distinct – digital and analog – types of control are balancing each other, such that upon limitations of one type of control (caused e.g. by mutation of a global DNA architectural factor) the other type can compensate for compromised regulation (Figure 6). This study thus paves the way towards a holistic theory of transcriptional regulation.

One prediction from the observed interdependence between digital and analog types of transcriptional control is that adaptive mutations in *E. coli* will affect the determinants of global DNA architecture. Indeed, a recent study of long-term experimental evolution in *E. coli* unmasking DNA topology as a key target for selection identified fitness-enhancing mutations in topoisomerase and *fis* genes [24]. Furthermore, such "evolved" populations possess high adaptational flexibility [25]. We propose that the buffering of transcriptional regulation by balancing effects of analog and digital types of control can counteract the reduction of adaptational flexibility caused by accumulation of mutations in bacteria [26]. In this respect it is revealing, that *fis* is a relatively late acquisition in bacterial evolution [27], whereas H-NS is implicated in regulating "adaptive" gene rearrangements and minimizing the cost of competitive fitness during horizontal gene transfer [16,28].

**Materials and Methods**

**Microarray and network data.** Transcript profiling for wild type, *fis* and *hns* LZ strains was carried out using *E. coli K12 V2 OciChip*™ DNA microarray. Each experiment was performed as two biological replicates with two technical replicates each, resulting in 28 cDNA microarray hybridisations. Scanned array images were quantified and normalized using the TM4 software package [29]. A one-class t-test was applied to replicated experiments to obtain affected genes with significant P-values (P < 0.05). DNA microarray data sets have been deposited in the Array Express data bank with the accession number E-TABM-86. For detailed DNA microarray data description and analyses see [8].

The latest version of the RegulonDB 5.6 data sets [1] "gene product" (http://regulondb.ccg.unam.mx/data/GeneProductSet.txt) and "regulatory network interactions" (http://regulondb.ccg.unam.mx/data/NetWorkSet.txt) were used for gene proximity network (GPN) and transcriptional regulatory network (TRN) generation, respectively.

**TRN construction.** Preceding the construction of effective TRNs, dimeric regulatory gene identifiers in the microarray data (*flhC, flhD; gatR_1, gatR_2; hupA, hupB; ihfA, ihfB; rcsA, rcsB*) were replaced by unique Regulon DB identifiers (*flhCflhD; gatR_1gatR_2; hupAhupB; ihfAihfB; rcsArcsB*). The effective TRN subnet of a DNA microarray transcript profile is the set of affected genes in the TRN and their regulatory interactions contained in RegulonDB (see Supporting information S1 for edge lists of the resulting effective TRNs). Connected components of an effective TRN emerge, if both regulating and regulated genes are affected in the transcript profile (see subnet analysis and Figure 2). Connected and unconnected subnet components were further analysed and can be found as edge lists in supporting information dataset S1.

**GPN construction.** Preceding GPN subnet construction, the inter-strain transcript profile data was split up into genes with positive and negative log ratios, respectively. Genes with positive log ratios refer to high transcript levels in wild type background, genes with negative log ratios refer to high transcript levels in *fis* or *hns* mutant background. GPN subnets of the split DNA microarray transcript profiles were generated based on genomic position of affected genes together with the proximity threshold $t$, given in in nucleotide bases (b). All affected genes with spatial distance (here distance is relating to ORF start and stop position) below the selected proximity threshold $t$ were considered as connected. GPN subnets were generated for a meaningful range of $1b < t < 10kb$, resulting in connected genes within an operon scale at $t \approx 10b$, up to completely conntected GPNs for $t > 10kb$. Connected and unconnected subnet components were further analysed and can be found as edge lists in supporting information dataset S2.

**Subnet analyses.** For each subnet, the control ratio $R$ was calculated as the number of connected nodes $N_{connected}$ (i.e. the size of the connected subnet component) over the number of isolated nodes $N_{isolated}$ (i.e. the size of the unconnected subnet component), $R = N_{connected} / N_{isolated}$. The control type confidence, CTC, is the z-score of $R$, calculated from the mean $R$ and its standard deviation obtained from 10000 runs of the corresponding null model. In the case of the digital null model, the same number of affected nodes was mapped randomly on the TRN (see Figure 2). For the analog null model, the same number of affected genes was mapped randomly on the positions in circular genome. The robustness of calculated ratios and CTCs was verified by 10% random data replacement with data of all affected genes from the remaining DNA microarray sets (see figure 1).

## Supporting Information

**Dataset S1.** Edge lists of the seven directed effective TRNs emerging from the analysis of the seven transcript profiles.

**Dataset S2.** Edge lists of the eight undirected GPNs emerging from the seperated analysis of the four inter-strain transcript profiles with positive and negative log ratio, respectively.

## Acknowledgements

**Funding.** CM was supported by a grant of the Darmstadt University of Technology. MG is supported by the DFG grant DFG-MU-2FIS.

**Author contribution.** CM, MG, MTH, and GM conceived the study. CM and MG analyzed the data. CM, MG, MTH, and GM wrote the paper.

## References

1. Salgado H, Gama-Castro S, Peralta-Gil M, Diaz-Peredo E, Sanchez-Solano F, et al. (2006) RegulonDB (version 5.0): Escherichia coli K-12 transcriptional regulatory network, operon organization, and growth conditions. Nucleic Acids Res 34:D394–397.
2. von Neumann J (1958) The computer and the brain. US: Yale University Press.
3. Gutierrez-Rios RM, Rosenblueth DA, Loza JA, Huerta AM, Glasner JD, et al. (2003) Regulatory Network of Escherichia coli: Consistency Between Literature Knowledge and Microarray Profiles. Genome Res 13:2435–2443.

4. Herrgard MJ, Covert MW, Palsson BO (2003) Reconciling Gene Expression Data With Known Genome-Scale Regulatory Network Structures. Genome Res 13:2423–2434.
5. Cheung KJ, Badarinarayana V, Selinger DW, Janse D, Church GM (2003) A microarray-based antibiotic screen identifies a regulatory role for supercoiling in the osmotic stress response of escherichia coli. Genome Res 13:206–215.
6. Jeong KS, Ahn J, Khodursky AB (2004) Spatial patterns of transcriptional activity in the chromosome of escherichia coli. Genome Biol 5:R86.
7. Peter B, Arsuaga J, Breier A, Khodursky A, Brown P, et al. (2004) Genomic transcriptional response to loss of chromosomal supercoiling in escherichia coli. Genome Biol 5:R87.
8. Blot N, Mavathur R, Geertz M, Travers A, Muskhelishvili G (2006) Homeostatic regulation of supercoiling sensitivity coordinates transcription of the bacterial genome. EMBO rep 7:710–715.
9. Allen TE, Price ND, Joyce AR, Palsson BØ (2006) Long-range periodic patterns in microbial genomes indicate significant multi-scale chromosomal organization. PLoS Comput Biol 2:e2.
10. Wright MA, Kharchenko P, Church GM, Segr`e D (2007) Chromosomal periodicity of evolutionarily conserved gene pairs. Proc Natl Acad Sci USA 104:10559–10564.
11. Shen-Orr SS, Milo R, Mangan S, Alon U (2002) Network motifs in the transcriptional regulation network of Escherichia coli. Nat Genet 31:64–68.
12. Yu H, Gerstein M (2006) Colloquium Papers: Genomic analysis of the hierarchical structure of regulatory networks. Proc Natl Acad Sci USA 103:14724–14731.
13. Travers A, Muskhelishvili G (2005) DNA supercoiling - a global transcriptional regulator for enterobacterial growth? Nat Rev Microbiol 3:157–169.
14. Dorman CJ (2004) H-NS: a universal regulator for a dynamic genome. Nature Rev Microbiol 2:391–400.
15. Grainger DC, Hurd D, Goldberg MD, Busby SJW (2006) Association of nucleoid proteins with coding and non-coding segments of the Escherichia coli genome. Nucleic Acids Res 34:4642–4652.
16. Dorman CJ (2007) H-NS, the genome sentinel. Nat Rev Microbiol 5:157–161.
17. Postow L, Hardy J Christine D and Arsuaga, Cozzarelli NR (2004) Topological domain structure of the Escherichia coli chromosome. Genes Dev 18:1766–1779.
18. Deng S, Stein R, Higgins N (2005) Organization of supercoil domains and their reorganization by transcription. Mol Microbiol 57:1511–21.
19. Cohen B, Mitra R, Hughes J, Church G (2000) A computational analysis of whole-genome expression data reveals chromosomal domains of gene expression. Nature Gen 26:183–186.
20. Kepes F (2004) Periodic Transcriptional Organization of the E. coli Genome. J Mol Bio 340:957–964.
21. Travers A, Muskhelishvili G (2005) Bacterial chromatin. Curr Opin Genet Dev 15:507–514.
22. Hardy CD, Cozzarelli NR (2005) A genetic selection for supercoiling mutants of escherichia coli reveals proteins implicated in chromosome structure. Mol Microbiol 57:1636–1652.
23. Luscombe NM, Babu MM, Yu H, Snyder M, Teichmann SA, et al. (2004) Genomic analysis of regulatory network dynamics reveals large topological changes. Nature 431:308–312.
24. Crozat E, Philippe N, Lenski RE, Geiselmann J, Schneider D (2005) Long-term

experimental evolution in escherichia coli. xii. dna topology as a key target of selection. Genetics 169:523–532.

25. Novak M, Pfeiffer T, Lenski RE, Sauer U, Bonhoeffer S (2006) Experimental tests for an evolutionary trade-off between growth rate and yield in E. coli. Am Nat 168:242–251.
26. Cooper V, Lenski R (2000) The population genetics of ecological specialization in evolving Escherichia coli populations. Nature 407:689–90
27. Morett E, Bork P (1998) Evolution of new protein function: recombinational enhancer Fis originated by horizontal gene transfer from the transcriptional regulator NtrC. FEBS Lett 433:108–12.
28. Gomez-Gomez J, Blazquez J, Baquero F, Martinez J (1997) H-NS and RpoS regulate emergence of Lac Ara+ mutants of Escherichia coli MCS2. J Bacteriol 179:4620–4622.
29 Saeed AI, Sharov V, White J, Li J, Liang W, et al. (2003) Tm4: a free, open-source system for microarray data management and analysis. Biotechniques 34:374–378.

**Abbreviations:** CTC, control type confidence, GPN, gene proximity network, TRN, transcriptional regulatory network.

**Figures**

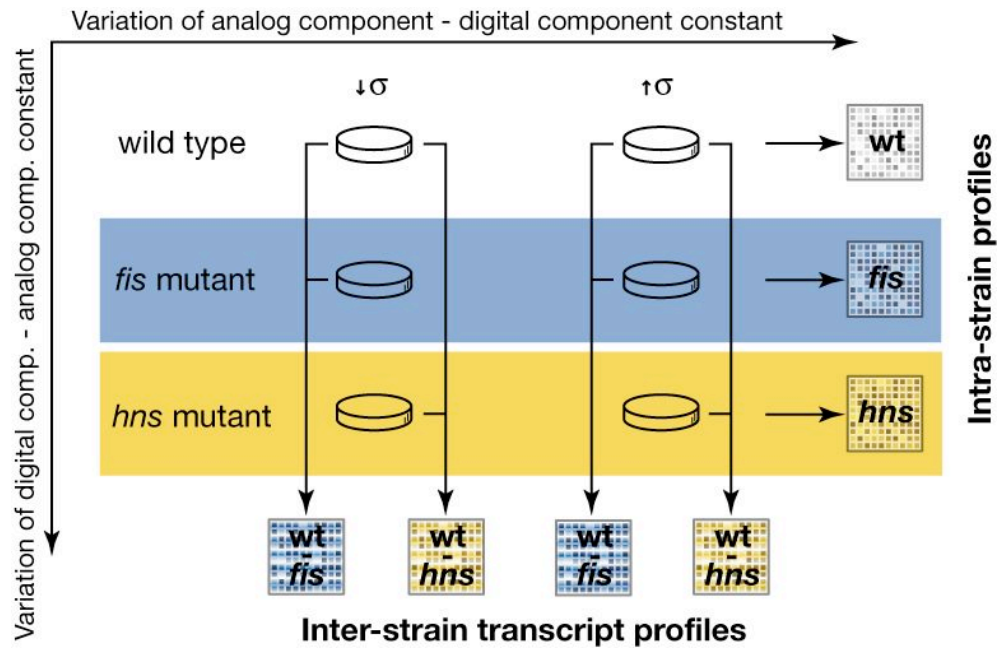

**Figure 1.** Experimental Setup.

In our experimental setup the transcript profiles of three *E. coli* strains (wild-type, *fis* mutant, *hns* mutant) are compared under low (< σ) and high superhelicity (< σ) and also with each other (vertical connections). The three intra-strain transcript profiles (**wt, *fis*, *hns***) show differentially expressed genes in response to variation of negative supercoiling but under a constant transcriptional regulatory network. The four inter-strain profiles (**wt-*fis*** and **wt-*hns*** for ↓σ and ↑σ each) show genes differentially expressed under constant supercoiling but with different genetic backgrounds. Note that alterations in superhelical density caused by mutations themselves are negligible compared to the experimentally induced changes of superhelicity [7].

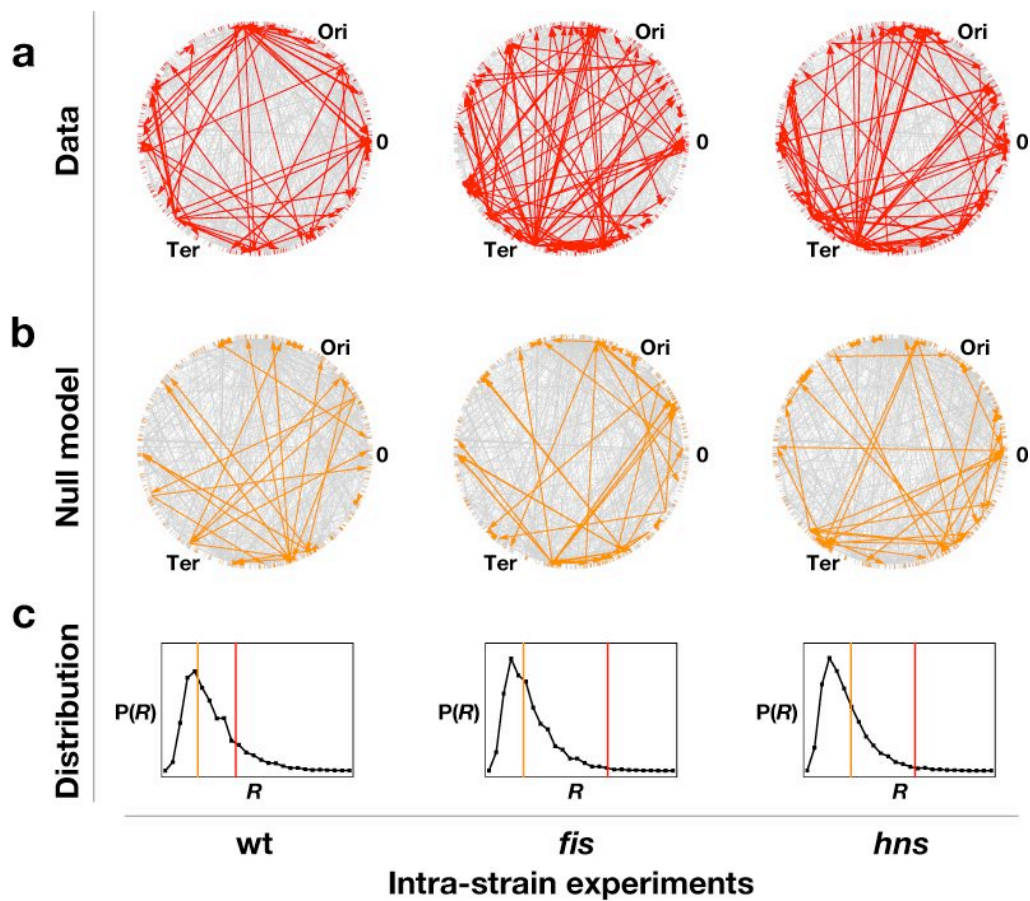

**Figure 2.** Calculation of the digital control type confidence (CTC).

(a) The effective TRNs (red) within the entire RegulonDB TRN (gray), mapped on the circular genome of *E. coli*. Only the three intra-strain experiments are shown.

(b) The effective TRNs (orange) within the entire RegulonDB TRN (gray) for a single null model realisation. The effective TRNs of the null models are less densely connected.

(c) The frequency distribution of *R* for 10,000 null models together with the actual values of *R* from the graphs shown in (a) and (b).

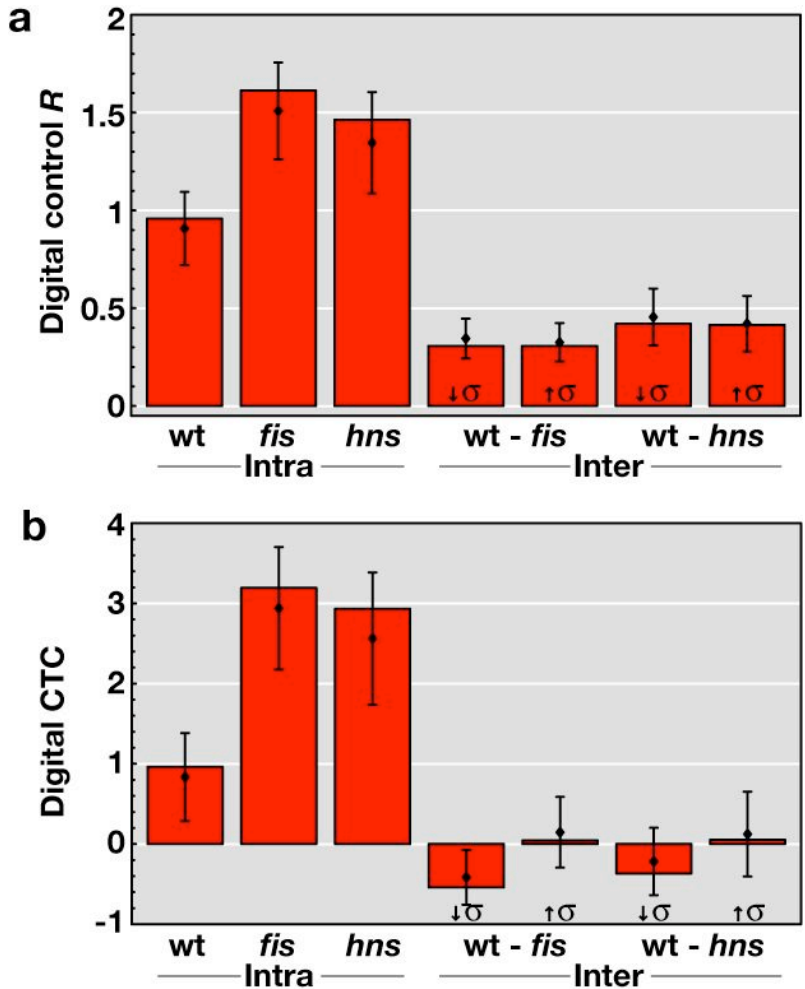

**Figure 3. Digital control *R* and control type confidence (CTC) of transcriptional regulation.**

(a) The digital control *R* is the number of connected nodes divided by the number of isolated nodes in the effective transcriptional regulatory networks of the three intra-strain experiments (**wt**, *fis*, *hns*) and the four inter-strain experiments (**wt-*fis*** and **wt-*hns*** for low (↓σ) and high (↑σ) negative supercoiling each).

(b) Digital CTC quantifies the deviation of the effective subnet based on significant expression changes from an appropriate null model. To estimate the sensitivity of the observables against noise, we replace 10% of all affected genes with randomly selected genes from the pool of affected genes in all other experiments. We then recalculate the digital control via the ratio *R* (a) and the corresponding CTC (b). We show the mean *R*s (diamonds) together with the standard deviation for 10,000 runs, and the mean CTCs (diamonds) together with the standard deviation for 1,000 runs, where the actual data is compared to 1,000 null model runs each.

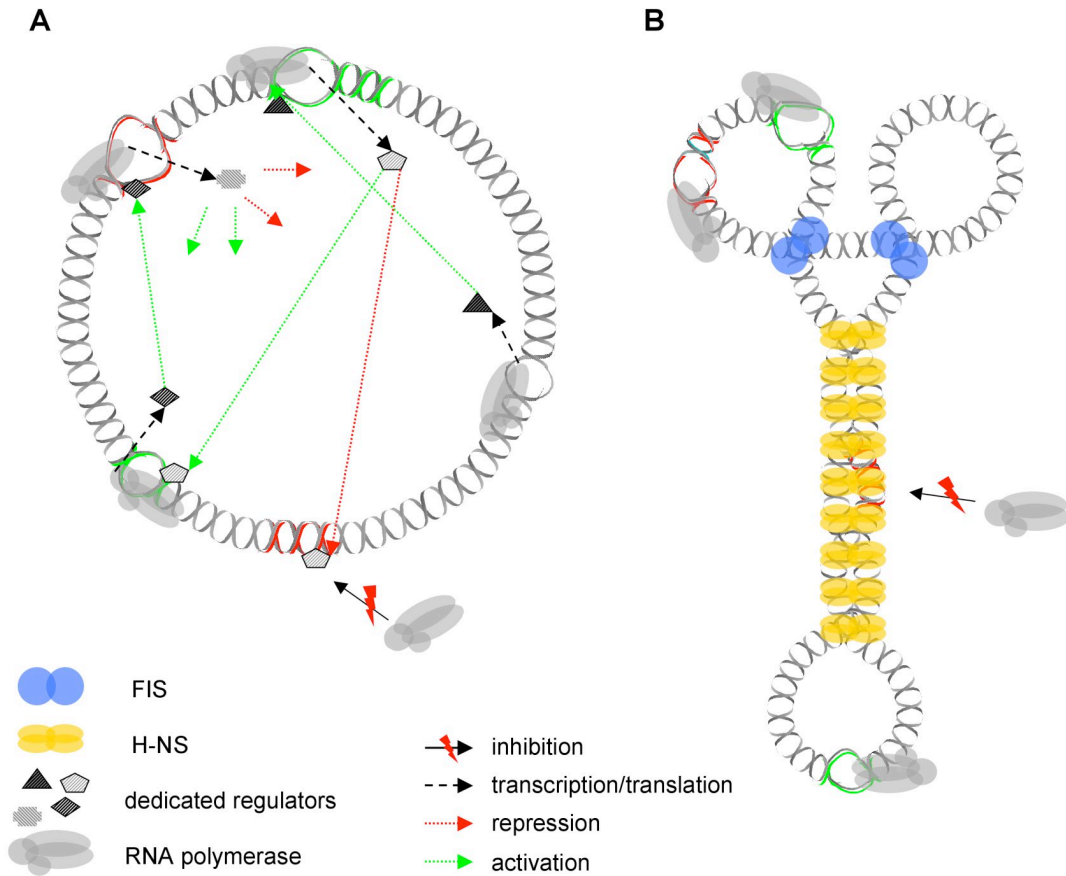

**Figure 4**. Schematic representation of digital-type vs. analog-type of regulation.

(A) Digital control: Dedicated regulators independently recruit polymerase to distantly located genes to either activate (green arrows) or repress (red arrows) their activity.

(B) Analog control: Abundant DNA architectural proteins (only FIS and H-NS are shown for simplicity) form topological domains, thus rendering the distant genes under independent digital control similarly accessible to polymerase. The activation of transcription is indicated by colored spheres associated with polymerase, repression of transcription by "red-flashed" arrows.

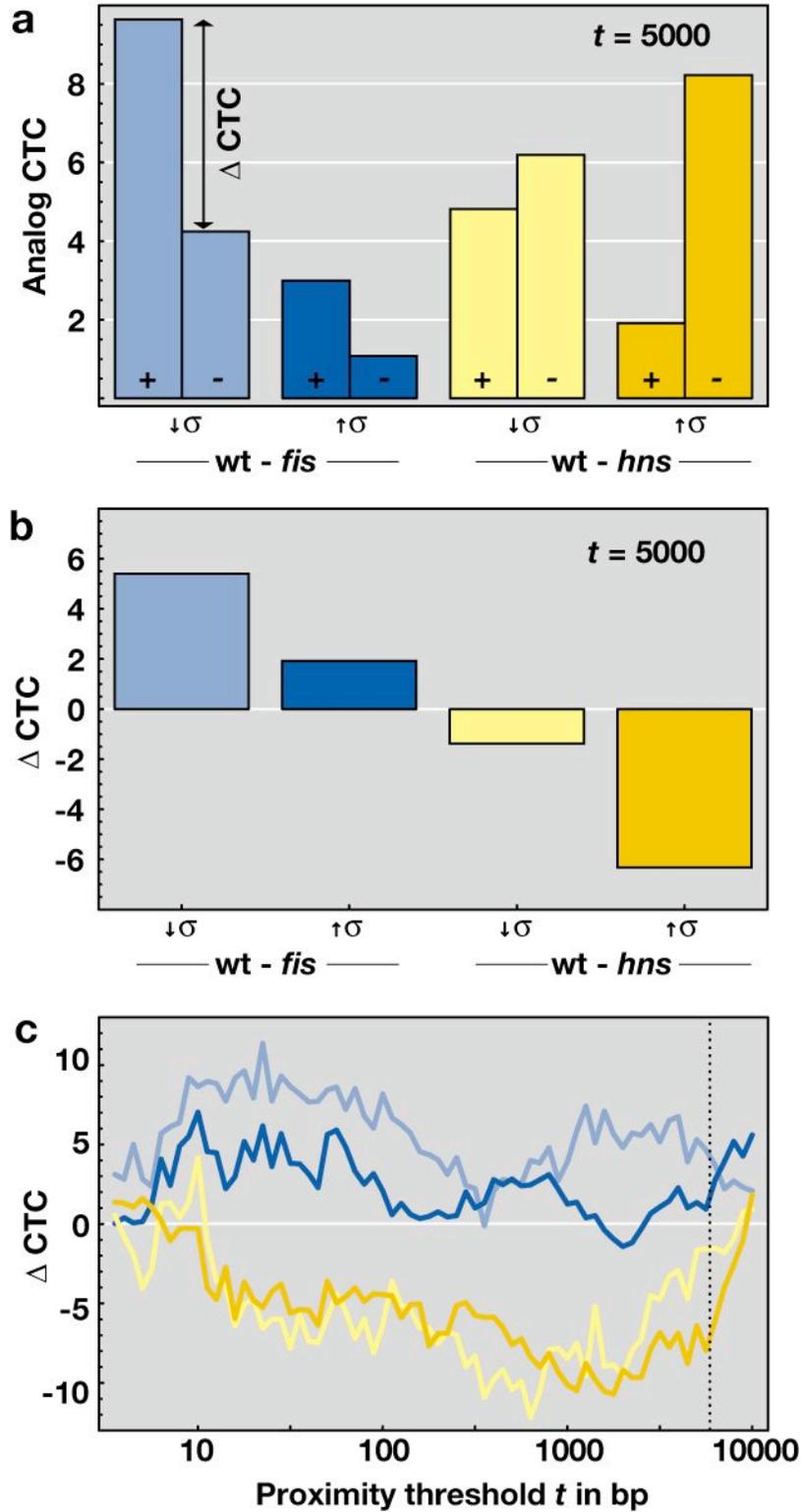

**Figure 5.** Analog control-type confidence of gene proximity networks.

(a) Analog CTC of the four inter-strain expression profiles at proximity threshold t = 5kb. The left (+) and right (-) bars correspond to expression data with log-ratios above and below 0, respectively. A positive log ratio (+) is associated with either a raised expression in wild type, or a inhibited expression in the mutant strain.

(b) Difference (left bar - right bar) of the analog CTCs from (a) for each inter-strain experiment at t = 5kb.

(c) Difference of the CTCs for each inter-strain experiment against the proximity threshold t. For t > 10kb, the effective GPNs are almost fully connected and a proper CTC calculation fails. Note that *fis* has a preponderantly activating and *hns* a preponderantly repressing regulatory effect.

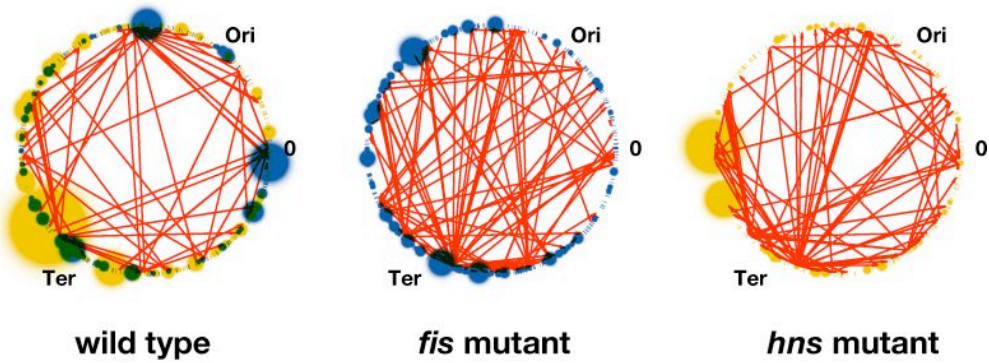

**Figure 6.** Distribution of control.

The organizational logic of transcriptional regulation revealed by combining the information on digital control (obtained from intra-strain experiments), and information on analog control (obtained from inter-strain experiments). Red arrows indicate the links of the effective TRNs. Colored segments on the circular genome are affected genes, as derived from the *fis* (blue) and *hns* (yellow) inter-strain experiments. The colored spheres indicate connected components in the effective GPN at a proximity threshold of t = 5kb. The actual size of the spheres is proportional to the diameter of the subnets spanned by each analyzed profile. The origin (Ori) and terminus (Ter) of chromosomal replication are also shown.